\documentclass[12pt]{JHEP3}

\newcommand{\bea}{\begin{eqnarray}}
\newcommand{\beal}[1]{\begin{eqnarray}\label{#1}}
\newcommand{\eea}{\end{eqnarray}} 
\newcommand{\be}{\begin{equation}} 
\newcommand{\bel}[1]{\begin{equation}\label{#1}}
\newcommand{\ee}{\end{equation}} 
\newcommand{\rf}[1]{(\ref{#1})}

\newcommand{\bit}{\begin{itemize}}
\newcommand{\eit}{\end{itemize}}
\newcommand{\ben}{\begin{enumerate}}
\newcommand{\een}{\end{enumerate}}

\newcommand{\sect}[2]{\section{#1}\label{#2}}

\def\half{\frac{1}{2}}
\def\third{\frac{1}{3}}

\def\alp{\leavevmode\ifmmode {\alpha^\prime} \else ${\alpha^\prime}$ \fi}
\def\mps{M_P^2}
\def\fnl{f_{NL}}
\def\nnl{n_{NL}}

\title{Inflation in DBI models with constant $\gamma$}

\preprint{}

\author{Micha\l\ Spali\'nski\footnote{Email: mspal@fuw.edu.pl}\\ So\l tan
  Institute for Nuclear Studies\\ ul. Ho\.za 69, 00-681 Warszawa, Polska.}

\abstract{Dirac-Born-Infeld scalar field theories which appear in the
  context of inflation in string theory in general have a field dependent
  speed of sound. It is however possible to write down DBI models which
  possess exact solutions characterized by a constant speed of sound
  different from unity. This requires that the potential and the 
  effective D-brane tension appearing in a DBI action have to be related in
  a specific way. This paper describes such models in general and presents
  some examples with a constant speed of sound $c_s<1$ for which the
  spectrum of scalar perturbations can be found analytically without
  resorting to the slow roll approximation.}

\keywords{Cosmology, Inflation, String Theory}

\begin{document}

\section{Introduction}

Nonlinear scalar field theories of the Dirac-Born-Infeld type 
\cite{Silverstein:2003hf}-\cite{Chen:2005fe} have
attracted much attention in recent years due to their role in models of
inflation based on string theory \cite{Dvali:1998pa,Kachru:2003sx}. 
These scenarios identify the inflaton
with the position of a mobile $D$-brane 
moving on a compact $6$-dimensional submanifold of spacetime (for reviews
and references see 
\cite{Cline:2006hu}-\cite{McAllister:2007bg}), which means that 
the inflaton is interpreted 
as an open string mode. This interpretation of the inflaton  
implies that that the effective field theory is rather distinct and 
very well motivated by string computations \cite{Leigh:1989jq}. 
Once all other degrees of freedom are eliminated the effective action in 
four dimensions is of the DBI type. This action involves a characteristic
square root reminiscent of the relativistic
Lorentz factor, and depends on a function, usually denoted by $f$, which
characterizes the local geometry of the compact manifold on which the
D-brane is moving. 
The ``Lorentz'' factor locally imposes a maximum speed on the
itinerant $D$-brane \cite{Silverstein:2003hf}.
When this factor is close to unity the action reduces to that of a
canonical scalar, but in general involves all powers of the inflaton
gradient.  

From a hydrodynamical point of view the presence of the field-dependent
``Lorentz'' factor $\gamma$ corresponds to a nontrivial, time dependent
speed of sound -- one finds \cite{Silverstein:2003hf} that
$c_s=1/\gamma$. Models of inflation with $c_s\neq 1$ were considered in a
more general setting in \cite{ArmendarizPicon:1999rj,Garriga:1999vw}. 
A small speed of sound has important observational consequences: it
translates into potentially observable levels of non-gaussianity in the CMB
spectrum. 

A non-constant speed of sound complicates the field equations considerably.
The object of this note is to point out that there is a special class of
DBI models which is in many ways as simple 
as models with a canonical kinetic energy term. These are models where the
potential and the function $f$ appearing in the ``Lorentz'' factor are
related in such a way as to admit solutions with constant $\gamma$ (in
general different from unity). These models have the constant $\gamma$
appearing as a parameter, whose deviation from unity measures the
deformation of the kinetic energy from the canonical form.  For such
``$\gamma$-deformed'' models the equations have the same form as in the
canonical case, apart from numerical factors involving $\gamma$.  In
particular, for potentials of interest for which a solution is available
when the kinetic terms take canonical form one can find the corresponding
family of solutions in terms of the constant $\gamma$.

For cosmology one is interested not only in the homogeneous limit, but also
in the perturbed system. The mode equation which determines the spectrum of
scalar perturbations is known also when the kinetic energy is
non-canonical \cite{Garriga:1999vw} and is of course more complicated than in the
canonical case.  
However when the speed of sound is constant it simplifies considerably, so
the study of perturbations in the class of models with constant $\gamma$ is
possible at the same level of approximation as required in the case of
$\gamma=1$. In particular, one can write down ``$\gamma$-deformed'' models
corresponding to known cases where the spectrum of scalar perturbations is
known analytically. The examples discussed are the constant potential (i.e. de
Sitter space), the exponential potential\footnote{This solution appears in
  the recent paper by Chimento and Lazkoz  \cite{Chimento:2007es} and is a
  special case of power law inflation in models of k-inflation
  \cite{ArmendarizPicon:1999rj,Garriga:1999vw}.} 
(leading to power law inflation) and the model introduced by Easther
\cite{Easther:1995pc}. These examples introduce $\gamma$ as a parameter in
addition to parameters present in the undeformed model. The deviation of
$\gamma$ form unity leads to non-gaussianity of the perturbation spectrum,
and the observational consequences of the deformation can be understood in
terms of commonly used observables $r$, $n_S$ and $\fnl$. In the example of
a constant potential the deformation parameter $\gamma$ turns out to be
very strongly constrained by the limits on the index of scalar
perturbations. In the case of an exponential potential the situation is
more interesting: the observables $r$, $n_S$ and $\fnl$ satisfy a relation
(described in section \ref{sectpow}), which could be tested observationally
by the Planck satellite experiment launching this year. 

This article is organized as follows: section \ref{dbisect} sets up some
notation and reviews the Hamilton-Jacobi formulation of DBI scalar field
theories. Section \ref{sectcons} presents the Hamilton-Jacobi equations for
DBI models 
with constant $\gamma$ and the following section discusses the computations
of the perturbation spectra.  Sections \ref{sectdes}, \ref{sectpow},
\ref{secteast} discuss the three examples where the perturbation spectrum
for the ``$\gamma$-deformed'' models is obtainable analytically.
In these examples the DBI scalar field theory is regarded as a
phenomenological model (in the spirit of \cite{Lidsey:2006ia}, for
example); they are not really motivated by string theory, apart from the
DBI kinetic terms. The effective brane tension $1/f$ is chosen so that a
given potential supports inflation with a constant speed of sound. From a
string theory perspective a more natural approach is to consider a throat
geometry which corresponds to a known string solution. An example of this 
sort is mentioned in the final section.

\sect{DBI scalar field theories}{dbisect}

DBI scalar field theories are an example of a wide class of scalar field
theories with non-canonical kinetic terms, whose significance in the
cosmological context was discussed in \cite{ArmendarizPicon:1999rj}. 
In models of D-brane inflation the effective action for the inflaton is
the Dirac-Born-Infeld action, which for spatially homogeneous inflaton
configurations takes the form \cite{Silverstein:2003hf} 
\bel{dbi}
S = - \int d^4x\ a(t)^3\ \Big(f(\phi)^{-1}(\sqrt{1-f(\phi) \dot{\phi}^2}-1) +  
V(\phi)\Big)  \ .
\ee
The function $f$ appearing here has the interpretation of inverse effective 
D-brane tension and can be expressed in terms of the warp
factor in the metric and the string length and coupling. 
Many aspects of DBI scalar field theories have recently been discussed 
in a number of papers \cite{Silverstein:2003hf}-\cite{Chen:2005fe}.

Einstein equations for homogeneous fields  reduce to
\bea
\dot{\rho} &=& - 3 H (p+\rho) \label{conserv}\\
3\mps H^2&=& \rho \label{friedman} \ ,
\eea
where $M_P$ is the reduced Planck mass ($M_P^2=1/8\pi G$), the dot indicates
a time derivative and $H\equiv \dot{a}/a$.
For the action \rf{dbi} the pressure and energy density are given by 
\bea
p     &=& \frac{\gamma-1}{f\gamma} - V(\phi) \label{pdbi}\\
\rho  &=& \frac{\gamma-1}{f} + V(\phi) \label{rhodbi}\ ,
\eea
where 
\bel{lorfac}
\gamma = \frac{1}{\sqrt{1-f(\phi)\dot{\phi}^2}} \ .
\ee
It is easily established that $\gamma$ is in fact the inverse speed of
sound in models of this type:
\be
c_{s}^{2}= \frac{\partial p}{\partial \dot \phi} / \frac{\partial
  \rho}{\partial \dot \phi} = \frac{1}{\gamma^{2}} \ .
\ee

The Hamilton-Jacobi form of the field equations which will be employed in
the following is obtained by eliminating explicit time dependence. Using 
\rf{conserv}--\rf{lorfac} one finds 
\bel{phidot}
\dot{\phi} = -\frac{2\mps}{\gamma} H'(\phi) \ .
\ee
This equation can easily be solved for $\dot{\phi}$ which allows $\gamma$
to be expressed as a function of $\phi$: 
\bel{gamma}
\gamma(\phi)=\sqrt{1+ 4 M_P^4 f(\phi) H'^2(\phi)} \ .
\ee
Using this in \rf{friedman} gives
\bel{hjdbi}
3\mps H^2(\phi) - V(\phi) =  \frac{\gamma(\phi)-1}{f(\phi)} \ .
\ee
This is the Hamilton-Jacobi equation for DBI
inflation \cite{Silverstein:2003hf,Spalinski:2007kt}. For comparison, the corresponding
equation for a canonical inflaton
\cite{Markov:1988yx}--\cite{Kinney:1997ne} is 
\bel{hjcan}
3 \mps H^2(\phi) - V(\phi) =  2 {M_P}^4 H'^2 (\phi) \ .
\ee
This of course follows from \rf{hjdbi} in the limit when $\gamma$ tends to
unity.

\sect{Models with constant $\gamma$}{sectcons}

As discussed in the introduction, the field dependence of $\gamma$ in
\rf{gamma} complicates subsequent analysis of DBI models significantly. 
It is 
natural to look for models where the speed of sound is 
constant \footnote{A model with a constant speed of sound different from
  unity is discussed in \cite{ArmendarizPicon:1999rj}.}. This is  
somewhat analogous to considering field theory models for which 
the ratio $w=p/\rho$ is constant (which leads to models of power law
inflation). This is not to suggest that such models are particularly likely
to emerge from string theory or that they are a priori especially suited to
describe the real world. They are however interesting as tractable models
which facilitate the exploration of cosmology with a low speed of sound. 

The requirement that $\gamma$ be constant imposes a relation
between the function $f$ and the potential 
$V$. One way to proceed is to 
set \rf{gamma} to a constant (denoted below by $\gamma$) so as to 
eliminate $f$ from \rf{hjcan}. This leads to 
\bel{redhj}
3\mps H^2(\phi) - V(\phi)  =  \frac{4M_P^4}{\gamma+1} H'(\phi)^2  \ .
\ee
This has the form of the Hamilton-Jacobi equation for a canonical inflaton 
\rf{hjcan} apart from the constant factor $1/(\gamma+1)$. This factor
cannot be eliminated, so this equation is not actually identical the one
valid in the case of a canonical scalar, but the form is the same. 
Thus, to write down a DBI model with a constant speed of sound for a given
potential requires solving \rf{redhj} and putting 
\bel{fsol}
f(\phi) = \frac{\gamma^2-1}{4M_P^4} \frac{1}{H'(\phi)^2} \ .
\ee
Alternatively, if one wanted to write down a model with a specific choice
of $f$, then \rf{fsol} could 
be solved for $H(\phi)$, which would then determine the required potential 
through \rf{redhj}. 

To illustrate these observations it is interesting to consider some
explicit examples, specifically solvable models known from studies of
inflation in the context of canonical scalar field theories.  As discussed
in the following section, when the speed of sound is constant the
computation of the scalar perturbations is technically the same as for the
case $c_s=1$. Thus it is interesting to consider examples where also the
scalar perturbation spectrum can be computed analytically -- this will be
the subject of sections \ref{sectdes}, \ref{sectpow} and \ref{secteast}.

\sect{Perturbation spectra}{sectpert}

The scalar perturbation spectrum for a general form of kinetic energy was
described by Garriga and Mukhanov \cite{Garriga:1999vw}. These results were
used by a number of authors to work out the spectrum for the special case
of DBI scalar field theories \cite{Alishahiha:2004eh,Shandera:2006ax}. The
perturbed 
system is parametrized in the longitudinal gauge as follows:
\be
\phi= \phi(t) + \delta\phi(x,t)
\ee
and 
\be
ds^{2} = -(1 + 2 \Psi)dt^{2} +a(t)^{2}(1 - 2 \Psi)dx^{2} \ .
\ee
The gauge invariant curvature perturbation is given by
\be
\zeta = \Psi + \frac{H}{\dot \phi} \delta\phi \ .
\ee
Introducing 
\bel{zdef}
z = \frac{a M_P}{c_s} \sqrt{2\epsilon}
\ee
and the scalar density perturbation $u=\zeta z$, the Fourier mode $u_{k}$
as a function of the wave number $k$ 
satisfies the equation \cite{Garriga:1999vw} 
\bel{modeq}
\frac{d^{2}u_{k}}{d\tau^{2}} + \left(\frac{k^2}{\gamma^2} -
\frac{1}{z} \frac{d^{2}z}{d\tau^{2}} \right) u_{k}=0 \ ,
\ee
where $\tau$ is the conformal 
time defined by $d\tau = dt/a$. 

For DBI models it follows that \cite{Alishahiha:2004eh} 
\bel{zdbi}
z = \frac{a}{H}  \gamma^{3/2} {\dot \phi} 
\ee
and then one can show that \cite{Shandera:2006ax}
\beal{zttz}
\frac{z_{,\tau\tau}}{z} &=& a^2 H^2 \Big((1+\epsilon-\eta
-\sigma/2)(2-\eta-\sigma/2) + \epsilon(2\epsilon-2\eta+\sigma) 
+ \\ \nonumber
&-& \eta (\epsilon+\sigma) + \xi - 
\half \sigma(2\sigma + \epsilon - \eta) + \half \epsilon\omega 
\Big) \ ,
\eea
where the Hubble slow roll parameters are 
\beal{srparams}
\epsilon &=& \frac{2\mps}{\gamma} (\frac{H'}{H})^2 \label{epsdef}\\
\eta &=& \frac{2\mps}{\gamma} \frac{H''}{H} \label{etadef}\\
\xi &=& \frac{4M_p^4}{\gamma^2} \left(\frac{H^{\prime}(\phi)
  H^{\prime\prime\prime}(\phi)}{H^2(\phi)} \right)  \\ 
\sigma &=& \frac{2\mps}{\gamma} \frac{H'}{H} \frac{\gamma'}{\gamma} \\ 
\omega &=& \frac{2M_p^2}{\gamma} \left(
\frac{\gamma^{\prime\prime}}{\gamma} \right) \ .
\eea

The above equations are valid in any DBI scalar field theory. In models
with constant $\gamma$ they simplify considerably, since in that case
$\sigma=\omega=0$. The dependence on $\gamma$ enters only via the $k^2$
term in the mode equation \rf{modeq} and implicitly via the slow roll
parameters appearing in equation \rf{zttz}, which reduces to the same form
as in models with a  canonical inflaton: 
\bel{zttcon}
\frac{z_{,\tau\tau}}{z} = a^2 H^2\left( (1+\epsilon-\eta) (2-\eta) +
2\epsilon^2 - 3\epsilon\eta +\xi\right) \ .
\ee
In general solving the mode equation still requires the slow roll
approximation, but the deviation of the speed of sound from unity does not
introduce any new complications. 

As is well known, non-canonical kinetic terms may
lead to 
non-gaussianities in the perturbation spectra. Adopting the frequently 
used estimator $f_{NL}$ as a rough measure, one may 
use the results of \cite{Alishahiha:2004eh,Chen:2006nt}
\bel{fnl}
f_{NL} = \frac{35}{108} (\gamma^2 - 1)
\ee
to interpret the constant value of $\gamma$ in terms of $\fnl$. 
Current observational limits allow $\fnl < 332$
\cite{Creminelli:2006rz}. This implies that $\gamma < 32$; since other
observables also depend on $\gamma$, in specific models stronger
constraints may appear, as discussed in the examples which follow.

\sect{The case of a constant potential}{sectdes}

The simplest example is a constant potential, which in the case of
canonical kinetic terms leads to exponential (de Sitter) expansion. In a
brane inflation context this might be relevant in a regime where the
dominant contribution to the potential comes from the brane/anti-brane
tension. The issue of terminating inflation (which would go on indefinitely
with constant energy density) is resolved in this setting due to the 
tachyon which appears in the spectrum when the D-brane gets within a string
length of the
anti-D-brane.  

For a constant potential $V(\phi)=V_0$ the Hamilton-Jacobi equation
\rf{redhj} reads 
\be
3\mps H^2(\phi) - V_0  =  \frac{4M_P^4}{\gamma+1} H'(\phi)^2  \ .
\ee
There are two solutions for $H$. The first is just $H=\sqrt{V_0/3\mps}$,
which  
describes exponential inflation (where the inflaton does not change with
time). This solution is valid for any function 
$f$; it follows from \rf{gamma} that in this case $c_s=\gamma=1$. 

There is also another solution, which for the case of a canonical
scalar was discussed previously by
Kinney \cite{Kinney:2005vj} (see also \cite{Spalinski:2007ef}). This
solution is given by  
\be
H(\phi) = \sqrt{\frac{V_0}{3\mps}} \cosh (\frac{\sqrt{3(\gamma+1)}}{2}
\frac{\phi}{M_P}) \ .
\ee
It describes a scalar which evolves with time according to 
\be
\dot{\phi} = -\frac{\sqrt{(\gamma+1) V_0}}{\gamma} \sinh
(\frac{\sqrt{3(\gamma+1)}}{2} \frac{\phi}{M_P}) \ .
\ee
When the scalar field reaches $\phi=0$ it stops and so the late-time limit
is de Sitter space with $H=\sqrt{V_0/3\mps}$ (independently of the value of
$\gamma$). This solution can be viewed as describing a transient stage in
the evolution of the system before the de Sitter
attractor is reached\footnote{Similar  
issues are addressed in \cite{Starobinsky:2005ab,Tzirakis:2007bf}.}.  

The Hubble slow roll parameters are 
\bea
\epsilon &=& \frac{3}{2}\frac{\gamma+1}{\gamma} \tanh^2
(\frac{\sqrt{3(\gamma+1)}}{2} \frac{\phi}{M_P}) \label{coneps} \\
\eta &=&  \frac{3}{2}\frac{\gamma+1}{\gamma}  \label{coneta} \\
\xi &=& 3\epsilon\label{conxi} \ .
\eea
In the canonical case (i.e. $\gamma=1$) one has $\eta=3$; this model
was discussed in \cite{Kinney:2005vj} 
as model of non-slow roll inflation with an exactly scale invariant
spectrum of scalar perturbations. The DBI generalization discussed here is
a family of models parametrized by $\gamma$, which remain non-slow roll
even for large $\gamma$: the slow roll parameters \rf{coneps}, \rf{coneta}
decrease with 
increasing $\gamma$, but the most one gets is a factor of $1/2$ relative
to the canonical result. These models provide a family of
examples for which the horizon-crossing formula for the power spectrum
breaks down in the same way as in the  $\gamma=1$ case discussed in
\cite{Kinney:2005vj}.  

Since $\epsilon\ll\eta$ in this case, one can simplify \rf{zttcon} by
dropping $\epsilon$-dependent terms. Introducing a new variable $y=k/aH$, 
the mode equation can be written as
\be
y^2\frac{d^2u_k}{dy^2} + [y^2 - (2 - 3\eta+\eta^2)] u_k = 0 \ .
\ee
Imposing the Bunch-Davies initial conditions in the standard way one finds
that the solution can be written in terms of a Bessel function
\be
u_k\sim y^{1/2} H_\nu(y) \ ,
\ee
where 
\be
\nu=|\frac{3}{2} - \eta| = \frac{3}{2\gamma} \ . 
\ee
Since in this example $\eta$ is large one has evaluate the spectrum of
scalar perturbations 
with care. As explained in \cite{Kinney:2005vj} the correct way to proceed
in this case is to define the power spectrum by 
\be
{\mathcal P}_S(k) = |\frac{u_k}{z}|^2 
\ee
in the limit of $y\rightarrow 0$. This leads to
\be
{\mathcal P}_S(k) \sim \frac{H^2}{\epsilon} y^{3(1-1/\gamma)} \ ,
\ee
so that that the scalar index in this case is given by
\bel{cpns}
n_S - 1 = \frac{d\ln({\mathcal P}_S)}{d\ln(k)}{\bigg|}_{aH=const} = 3
(1-\frac{1}{\gamma}) \ .
\ee
When $\gamma=1$ the spectrum is scale invariant (as shown already in
\cite{Kinney:2005vj}), but taking $\gamma>1$ yields a blue spectrum (and no 
running). The 
WMAP3 analysis \cite{Spergel:2006hy} favours $n_S - 1 \approx 0.96$ if
negligible tensor 
perturbations are assumed. Relaxing this assumption, as well as allowing
contributions from cosmic strings favours larger values 
of $n_S$ \cite{Bevis:2007gh}.  
Even if $n_S$ is allowed to be somewhat larger than unity,
\rf{cpns} implies that $\gamma$ has to be quite close to the canonical
value. To linear order in $n_S-1$ one has
\bel{gamsol}
\gamma \approx 1+\third (n_S-1) \ .
\ee
This can be phrased in terms of non-gaussianity using \rf{gamsol}, which
leads to  
\be
f_{NL} \approx \frac{1}{5} (n_S - 1) \ .
\ee
Observational limits on $n_S$ imply that this model gives negligibly small
non-gaussianity. 
Thus, while formally one can consider 
any $\gamma\geq 1$, observation severely limits the magnitude of such a 
deformation in this case.

\sect{Power law inflation}{sectpow}

The second example is the model of power law inflation
\cite{Lucchin:1984yf}, 
characterized by an exponential potential:
\bel{potexp}
V(\phi) = V_0 \exp(2 b \frac{\phi}{M_P}) \ ,
\ee
where $b$ is a constant parameter. In the case of a canonical scalar it
leads to power law inflation when $b<1/\sqrt{2}$. In a D-brane inflation
context, as mentioned in the previous section, inflation terminates 
because of the appearance of a 
tachyon the D-brane gets close to the anti-D-brane.  

To find a constant
$\gamma$ DBI model with this potential one has to solve \rf{redhj}. The
obvious solution is 
\be
H(\phi) = H_0 \exp (b  \frac{\phi}{M_P}) \ ,
\ee
with
\be
H_0^2 = \frac{V_0}{3\mps (1-\frac{4b^2}{3(\gamma+1)})} \ .
\ee
The form of the function $f$ required for constant $\gamma$ follows from
\rf{gamma}: 
\be
f(\phi) = \frac{\gamma-1}{V_0} (\frac{3}{4} \frac{\gamma+1}{b^2} - 1)
\exp(-2 b  \frac{\phi}{M_P}) \ .
\ee
This solution still describes power law inflation, as was recently
discussed by Chimento and Lazkoz \cite{Chimento:2007es}. 

In general, power
law inflation in DBI models requires 
\cite{Spalinski:2007dv} 
\bel{bardbi}
4\mps H'^2 = 3 (w+1) H^2 \gamma(\phi) \ .
\ee
This follows from \rf{pdbi} and \rf{rhodbi} which imply
\be
p+\rho=\frac{\gamma^2(\phi)-1}{f\gamma(\phi)}\ .
\ee
Equating this to $(w+1)\rho$ and using \rf{gamma} and \rf{hjdbi} gives
\rf{bardbi}. For the model discussed in this section \rf{bardbi} implies
that
\be
w = \frac{4b^2}{3\gamma} - 1 \ ,
\ee
which is indeed constant. 

For the present model the Hubble slow roll parameters are 
\bel{epspow}
\epsilon = \frac{2 b^2}{\gamma} 
\ee
and $\eta=\epsilon$. Inflation occurs when $\epsilon<1$, which means 
\be
b<\sqrt{\frac{\gamma}{2}} \ ,
\ee
so even a steep exponential potential can be suitable for inflation if
the constant $\gamma$ is taken large enough. 

The spectrum of scalar perturbations in this model is
computable exactly in the same way as in the canonical case. Since the
mode equation can be solved analytically in the case of power 
law inflation, the present model provides a solvable example with
$c_s\neq 1$. 
The calculation proceeds in the same way as the usual calculation for power
law inflation \cite{Lyth:1991bc}, except that now $\epsilon$ 
depends on $\gamma$ as seen from \rf{epspow}, not just on the parameter $b$ 
of the potential.  
The resulting scalar spectral index which follows is given by  
\be
n_S -1 \equiv  \frac{d\ln{\mathcal P}_S}{d\ln\:k} 
\simeq - 2\epsilon (1+\epsilon) \label{plns} \ .
\ee
There is no running of the scalar index in this model, as always for power
law inflation. 

The tensor mode spectral density is given by
\be\label{tensor}
{\mathcal P}_T=\left.\frac{2H^2}{M_p^2\pi^2} \right|_{k=aH} \ ,
\ee
so the ratio of
power in tensor modes versus scalar modes is given exactly by 
\be
\label{tsratio}
r=\frac{16\epsilon}{\gamma} \ .
\ee
The tensor index is 
\bel{plnt}
n_T = -2\epsilon \ .
\ee

The non-gaussianity in this model can be described in terms of the estimator
$f_{NL}$ given by \rf{fnl}. The discussion can be framed in terms of the
consistency relation for DBI models obtained by Lidsey and Seery
\cite{Lidsey:2006ia}:
\bel{lidseer}
8 n_T = -r\sqrt{1+\frac{108}{35} f_{NL}}
\ee
and the relation described in \cite{Spalinski:2007qy}:
\bel{powercons}
n_S - n_T = 1 + \half \nnl (1+\frac{35}{108}\frac{1}{\fnl})^{-1} \ ,
\ee
where 
\bel{nnl}
\nnl \equiv \frac{d\ln\fnl}{d\ln k}
\ee
measures the running of the non-gaussianity estimator \cite{Chen:2005fe}. 
Since $\gamma$ is constant, so is $f_{NL}$, and therefore
$n_{NL}=0$. It follows that in this
case $n_S - 1 = n_T$ independently of $\gamma$, just as in the case of
canonical power law inflation. This together with 
\rf{lidseer} implies that  
\bel{gamls}
8 (n_S - 1) = - r \sqrt{1 + \frac{108}{35} f_{NL}} \ .
\ee
This is a relation between observable parameters, and it is quite possible
that it may be tested in the foreseeable future. 

One way the relation \rf{gamls} can be satisfied is exact scale invariance:
$n_S=1$ with no tensor perturbations, $r=0$. In such a situation this model does
not constrain 
$\gamma$ beyond what follows from existing bounds on
non-gaussianity. i.e. $\gamma<32$.  
On the other hand, if one assumes $r\neq 0$, one can rewrite \rf{gamls} as  
\bel{fnlpli}
\fnl = \frac{35}{108} \Big(\frac{64 (n_S -1 )^2}{r^2} - 1 \Big) \ ,
\ee
which shows that unless the model is almost 
exactly scale invariant one has significant non-gaussianity for small
$r$. For example, if the Planck satellite experiment observes $r$ close to 
the projected experimental sensitivity $r\approx 0.01$, equation
\rf{fnlpli} with $n_S=0.96$ implies that $\fnl$ in this model is of the
order of the existing observational bound ($\fnl\approx 332$). 

One can also ask what level of tensor perturbations is to be expected from
\rf{gamls}; assuming $n_S\approx 0.96$ one finds $r=0.32$ for $\fnl=0$,
dropping to $0.01$ for $\fnl=332$ (the current limit
\cite{Creminelli:2006rz}). 

The fact that this model is solvable is made less surprising by the
fact\footnote{I thank the anonymous referee for bringing this to my
  attention.} that by introducing a new field variable $\psi =
\exp(-b\phi)$ one can write the Lagrangian in the form \be p
= \frac{g(X)}{\psi^2} \ee where $X=\half\dot{\psi}^2$. Lagrangians of this
form were first considered in \cite{ArmendarizPicon:1999rj}, where it was
pointed out that they lead to power law inflation in the context of general
k-inflation models.  The perturbation spectra for all models of this type
were subsequently found by Garriga and Mukhanov \cite{Garriga:1999vw}.

This example is interesting and perhaps important not only because it
is a solvable 
deformation of canonical power law inflation, but it may have some
phenomenological 
appeal, as it shows that taking $\gamma>1$ can make inflation possible even
when the 
potential is too steep too give inflation at $\gamma=1$.

\sect{A DBI analog of Easther's model}{secteast}

Easther has set up another model which for which the Mukhanov equation is
exactly solvable \cite{Easther:1995pc}. This model is phenomenologically
unappealing, since the spectrum of scalar perturbations which follows from
it is far from what is observed. However, since the model is solvable it
provides another illustration for the main subject of this paper, namely
inflation in DBI models with constant speed of sound. 
A priori this type of
deformation of Easther's
model could lead to more interesting perturbation spectra. Unfortunately,  
from a physical perspective this 
deformation turns out not to be an improvement. 

Easther's observation is that the Mukhanov
equation reduces to the harmonic oscillator equation if $z$ is independent
of the scalar field. This gives a differential equation for $H(\phi)$ which
is solvable analytically and the corresponding potential can
be identified using the 
Hamilton-Jacobi equation. This scheme can also be carried out in the
present context. Using \rf{zdbi}, a simple calculation shows that 
\be
\frac{dz}{d\phi} = -\frac{2}{3} a \gamma^{1/2} (-\frac{\gamma}{2\mps} +
\frac{H''}{H} - (\frac{H'}{H})^2) \ .
\ee
Setting this to zero gives a differential equation for  $H(\phi)$ which is
of the 
same form as Easther's equation, apart from the factor of $\gamma$ in the
first term:
\be
-\frac{\gamma}{2\mps} + \frac{H''}{H} - (\frac{H'}{H})^2 = 0 \ .
\ee
The solution is given by
\be
H(\phi) = H_0 \exp (\frac{1}{4} \gamma (\frac{\phi}{M_P})^2 + \beta\phi) \ ,
\ee
where $\beta, H_0$ are integration constants. Since $\beta$ can be eliminated by
shifting $\phi$ it can be set to zero without loss of generality. 
The potential and the function $f$
can then be determined from \rf{gamma} and \rf{redhj}: 
\bea
V(\phi) &=&  3\mps H_0^2 \left(1-\frac{\gamma^2}{\gamma+1}
\frac{\phi^2}{3\mps}\right) \exp (\frac{1}{2}\gamma(\frac{\phi}{M_P})^2) \\ 
f(\phi) &=& \frac{\gamma^2-1}{\gamma^2} \frac{1}{H_0^2\phi^2}
\exp(-\frac{1}{2}\gamma(\frac{\phi}{M_P})^2) \ .
\eea
The slow roll parameter $\epsilon$ is given by
\be
\epsilon = \frac{1}{6}\gamma(\frac{\phi}{M_P})^2 \ .
\ee
Note that in this example increasing
$\gamma$ acts to shrink the range of $\phi$
where inflation takes place. 
 
The power spectrum of scalar perturbations comes out as
\be
P = \frac{3}{16\pi^2\mps} \gamma \frac{H^2}{H'} \ ,
\ee
evaluated at $k=\gamma a H$. A simple calculation then shows that the
scalar index in this model is independent of $\gamma$, and given by
Easther's result $n_s=3$. Thus this model remains outside the limits set by
observation even when deformed as discussed above.

\sect{Closing remarks}{sectconc}

DBI scalar field theories follow from the interpretation of the inflaton as
an open string mode. Despite difficulties in embedding them in a fully
satisfactory way in string theory\cite{Baumann:2006cd} it is very important
to understand their dynamics.

This note described a special class of 
DBI models with an arbitrary constant speed of sound $c_s<1$. 
The three examples of this kind which were discussed in the text provide
examples of inflationary scalar field theories for which the spectrum of
primordial 
scalar perturbations can be computed analytically (at the level of linear
perturbation theory, but without appeal to any slow-roll approximation).  
The existence of exact analytic results can be useful as a test of
approximation schemes and numerical calculations. 
From a phenomenological perspective it may be interesting that
``$\gamma$-deformed'' models with exponential potentials inflate even when
the potential is too steep for inflation with canonical kinetic energy.
These models also lead an interesting relation between observable
parameters (eq. \rf{fnlpli}), which may be tested observationally by the
Planck satellite. 

In the last three sections the potential was assumed and the throat
function $f$ was chosen so as to support evolution with a constant speed of
sound. 
In a string theory setting
it may be more natural to proceed in the opposite fashion: choose a
``throat function'' $f$ corresponding to a consistent string
compactification and look at models with potentials chosen to satisfy
\rf{redhj}. One may for example inquire what is the potential which
leads 
to inflation with a constant speed of sound in the case of the AdS throat
frequently used as an approximation of the Klebanov-Strassler geometry away
from the tip. Using $f(\phi)=\lambda/\phi^4$ in \rf{fsol} shows that to
have constant $\gamma$ one needs 
\bel{hads} H(\phi) = \frac{1}{6\mps}
\sqrt{\frac{\gamma^2-1}{\lambda}} \phi^3 
\ee 
plus a constant, which will be
set to zero for the moment. Inflation takes place when 
\be \epsilon =
\frac{18}{\gamma} (\frac{M_P}{\phi})^2 
\ee 
is less than one. One also finds
$\eta = 2/3\epsilon$ in this model.  The potential which follows from
\rf{hads} via \rf{redhj} is 
\be V = \frac{\gamma-1}{\lambda}
\Big(\frac{\gamma+1}{12\mps} \phi^6 - \phi^4\Big) 
\ee 
It is easy to see
that when $\epsilon \ll 1$, the $\phi^6$ term dominates the
potential. Allowing for a constant term in \rf{hads} implies that other
powers of $\phi$ also appear, apart from a mass term for
$\phi$. Having constant $\gamma$ is therefore possible for the AdS throat
at the cost of fine tuning the potential; in particular, for exactly
constant $\gamma$ there can be no $\phi^2$ term.

{\em Note added:} a few weeks after this article was posted on arXiv.org an
interesting paper \cite{Kinney:2007ag} appeared which explores models of
this type as well as their generalization in the spirit of
\cite{Spalinski:2007ef}.

%\newpage

\end{document}